\documentclass[sigconf]{acmart}
\usepackage{multirow}

\def\BibTeX{{\rm B\kern-.05em{\sc i\kern-.025em b}\kern-.08emT\kern-.1667em\lower.7ex\hbox{E}\kern-.125emX}}

\acmBooktitle{} 
\acmPrice{}
\acmISBN{}

\begin{document}

%%
%% The "title" command has an optional parameter,
%% allowing the author to define a "short title" to be used in page headers.
\title{Item Matching using Text Description and Similarity Search }

%%
%% The "author" command and its associated commands are used to define
%% the authors and their affiliations.
%% Of note is the shared affiliation of the first two authors, and the
%% "authornote" and "authornotemark" commands
%% used to denote shared contribution to the research.

\author{Ana Paula Appel}
\affiliation{%
   \institution{IBM}
%   \streetaddress{1 Th{\o}rv{\"a}ld Circle}
%   \city{Hekla}
   \country{Brazil}}
 \email{apappel@br.ibm.com}
 
  \author{Anderson Luis de Paula Silva}
\affiliation{%
   \institution{IBM}
%   \streetaddress{1 Th{\o}rv{\"a}ld Circle}
%   \city{Hekla}
   \country{Brazil}}
 \email{anderson.silva@br.ibm.com}

\author{Adriana Reigota Silva}
\affiliation{%
   \institution{IBM}
%   \streetaddress{1 Th{\o}rv{\"a}ld Circle}
%   \city{Hekla}
   \country{Brazil}}
 \email{adrianareigota@ibm.com}
 
\author{Caique Dutra Santos}
\affiliation{%
   \institution{IBM}
%   \streetaddress{1 Th{\o}rv{\"a}ld Circle}
%   \city{Hekla}
   \country{Brazil}}
 \email{caique.santos@ibm.com}
 
\author{Thiago Lobo da Silva}
\affiliation{%
   \institution{IBM}
%   \streetaddress{1 Th{\o}rv{\"a}ld Circle}
%   \city{Hekla}
   \country{Brazil}}
 \email{thiagosilva@ibm.com}
 
\author{Rafael Poggi de Araujo}
\affiliation{%
   \institution{IBM}
%   \streetaddress{1 Th{\o}rv{\"a}ld Circle}
%   \city{Hekla}
   \country{Brazil}}
 \email{Rafael.Poggi@ibm.com}

\author{Luiz Carlos Faray de Aquino}
\affiliation{%
   \institution{IBM}
%   \streetaddress{1 Th{\o}rv{\"a}ld Circle}
%   \city{Hekla}
   \country{Brazil}}
 \email{luiz.carlos.faray@ibm.com}

\renewcommand{\shortauthors}{Appel, et al.}

%%
%% The abstract is a short summary of the work to be presented in the
%% article.
\begin{abstract}
In this paper, we focus on the problem of item matching using only the description. Those specific item not only lack of a unique code but also contain short text description, making the item matching processes difficult. Our goal is comparing products using only the description provided by the purchase process. Therefore, evaluating other characteristics differences can uncover possible flaws during the acquiring phase. However, the text of the items that we were working on was very small, with numbers due to the nature of the products and we have a limited amount of time to develop the solution which was 8 weeks. As result, we showed that working using a well-oriented methodology we were able to deliver a successful MVP and achieve the results expected with up to 55\% match.

\end{abstract}

%%
%% Keywords. The author(s) should pick words that accurately describe
%% the work being presented. Separate the keywords with commas.
\keywords{similarity search, index, item matching}
\maketitle

\section{Introduction}

Item matching~\cite{lei2019cloth} is a core function in several domains. In online marketplace where retailers compare new and updated product information against existing listings to optimized customer experience~\cite{mohammadi2021smart}. List comparison where retailers compare their listings with competitors to identify differences in price and inventory. It is also important in others domain that work with small text as for example search for similar text in chatbot logs, past purchase, social media, etc. 

However our challenge is not only develop a solution that solves client problem but also we need to do this in a short amount of time and in a scalable way allowing our client be able to continue working in the solution adding new features. In order to do this we apply an approach where we co-create, co-execute and co-operate. 

In more detail, during the co-creation we put together a diverse team of subject matter experts immersed in intensive design thinking and research activities to expose the true nature of a client’s opportunity and establish alignment on a ``big idea'' to address a specific pain point for a typical end user and create a vision for a minimum viable product (MVP).

After that, in co-execute a solution development cycle that uses Agile and DevOps practices to quickly launch and test an MVP, while capitalizing on business and technology expertise. In this phase is important also have technical team of client working closely. Our squad is usually composed by developer, data scientist, cloud architect and designer. The goal of this phase is validate and improve the MVP’s value in the marketplace through an iterative process of testing, measuring and re-launching.

In the end, we delivery the MVP but we co-operate with clients and other business unit that will continue to harden and scale the solution, expanding DevOps practices to broaden feature sets, stress test code, strengthen security and resilience, deploy solutions widely, and expand capabilities, empowered with the confidence to continue to innovate and transform. 

In the following sections we will detail our technical work where we built this solution in 8 weeks and allowed the client to retrieve and compare the items using only a poor text description. We were able to match up to 55\% of the items in a scalable way without using any identifier. Our goal is identify similar items bought in the past in order to compare. 

\subsection{Related Work}

Search is a feature highly present in our life, since the start of relational database we are always want to search for things. Relational databases are with us for more than half a century and it works perfectly fine in search millions of objects. However, when we do not have a key in order to index our products it becomes a little more complicate. Similar search on documents have been with us around 30 years and drive the internet success, using efficient retrieval system and the right data structure to index billions of documents and allows to search in milliseconds.

In order to be able to index and search text we need to create a representation that allow us compare they efficiently, for example, use bag of words~\cite{robertson2004understanding}, TF-IDF~\cite{church1999inverse}, topic modeling~\cite{wallach2006topic}, etc.

Despite the fact that these techniques work pretty well and are scalable, they do not carry meaning and context with them and are too shallow. However, these techniques are the base from the development of better methods. 

The advent of embeddings, making use of small vectors to represent text - as item descriptions, allow us to build index that can help users find similar items as music, products, videos, recipes, etc. 

As we know text can be represent by words, sentences and documents and deep learning is a powerful tool to work with these forms and lenghts. 
One of the most popular embedding is Word2vec~\cite{church2017word2vec}, it was one of the first and most applied to text. With it is possible correlate words with other words based on the meaning. 

The evolution of word embedding are the transformers, that allows encode a whole sentence which makes possible use the relation between sentence word and the order among them. USE (Universal Sentence Encoding)~\cite{yang2019multilingual} is one that works well and has a multilingual version.
One special kind of transformer very popular is Bert~\cite{devlin2018bert} with also a multilingual version and the more recently GPT-3~\cite{brown2020language} and OPT-175~\cite{zhang2022opt} models allowing we go further with search and complete the whole sentence.

With the embeddings created we can finally index and search. We are able to do that using K-NN~\cite{fix1989discriminatory} (k nearest neighbor) query and Range query \cite{agarwal1999geometric}. Basically, K-NN computes the distance between the query point with others in the search space to find the $k$ similar points. This method suffer a lot evolution over the years in order to become efficiently as approximate k nearest neighbors algorithms \cite{hajebi2011fast}.

In order to use this in real life problems we need the right implementation of K-NN index~\cite{zhao2014locality}. Using quantization techniques we are able to make real index and search billions of vectors of embeddings using only few gigabytes of memory. Usually, to choose we look into three aspects: latency - time consuming to index return results, Recall - how is the retrieve of the queries, Memory - how much memory it is consuming. 

Depend of amount of data that one is trying to index and search we can choose to brute force but higher accuracy, in memory algorithms that are fast, quantization or disk index solutions. 

Currently we have some libraries that implements K-NN and Range searches over index of embeddings: 

\begin{itemize}
    \item FAISS~\cite{johnson2019billion} - One of the best library with a very nice interface and several methods to index and search. Also, allows add more content after the index creation.
    \item Hnswlib~\cite{malkov2018efficient} is the fastest implementation, highly specialized and optimized.
    \item Annoy~\cite{li2019approximate} is another K-NN library implemented by Spotify.
    \item Scann~\cite{avq_2020} use anisotropic quantization and is one of the faster index outperformed HNSW in speed and recall. 
    \item Catalyzer~\cite{sablayrolles2018spreading} use neural network to train quantizer.
\end{itemize}

Based on requirements as maturity level to deploy, efficiency, scalability, flexibility and possibility to add new items instead of re-build we decide to use FAISS in our solution.

%Artigos sobre MRO e busca por similaridade

%https://databricks.com/blog/2021/05/24/machine-learning-based-item-matching-for-retailers-and-brands.html

%https://rom1504.medium.com/semantic-search-with-embeddings-index-anything-8fb18556443c

\section{Problem Definition}

In item match process, the ability of identify correctly the items is very important. This process is trivial when the product in question has a identifier, that is, the problem reduce to traditional query on a relational database. However, for products that do not have a identifier this become a more complex retrieval information problem that requires the use of a specialized data structure in order to perform nearest neighbor and range queries. 

In a more formal way: 

\begin{definition}
Given a set of vectors $x_i$ with dimension $d$ represent a product description we need find all the past purchases where $i = argmin_i ||x - x_i||$ where $||.||$ is the Euclidean distance (L2).
\end{definition}

For our case we define that we will consider a match if the distance between the item and the closest neighbor ($k=1$) is less than $0.4$. This value was elicited after we make an evaluation of a sample of items with users that worked as a subject matter expert (SME) in the development of our solution (co-execute phase). This value will changed based on the dataset used. 

However, FAISS is not able to work with a tie list and we do not know how many items we have at the same distance of the $k=1$ neighbor. To solve this, we perform a K-NN and retrieve the closes point, if it fits our requirement of distance we perform a range query with the distance of this closest point plus a small number (0.00001) to retrieve all items that are at that distance. 

A range search returns all vectors within a radius around the query point (as opposed to the k nearest ones).

\section{Data and Solution}

Our dataset with around 301,204 items. In this dataset we had one third of orders (87.331) that were purchases without identifier and two thirds (213.873) purchases with identifier. 

\begin{figure}[h]
\caption{Text Length distribution of purchase description}
\centering
\includegraphics[width=0.45\textwidth]{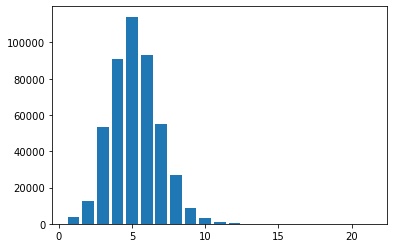}
\label{fig:distribution}
\end{figure}

As already said, the search based on product description was the focus of our solution. The text used to described the product was very short and poor, we present the distribution in Figure \ref{fig:distribution}. Most of the text are between 4 to 6 words without remove any stop words. Another problem was that because of the nature of the product we have a lot of numbers in this text description, for example: screwdriver 1/8 x 4'' and screwdriver 1/4 x 10'', are different and the price can vary significantly. 

In order to build our index we first generate the embeddings from text description using USE (universal sentence embedding), after that we create an index using FAISS, we use \textit{IndexFlatL2} that allows us to do exact search and since we are indexing less than half a million and it fits on memory this is the best choice for our solution. The rest of the data we store in a traditional database and the index of each tuple is the same of the embeddings on our FAISS index. 

After the index construction, we are able to use the index to query the products. One challenge is that we don't know neither how many neighbors we should query or the distance we should use, because for each product we have different number of similar products with different distance. However we know that the similarity will be higher since the text is short and any variation may indicate a different product. Before set up our strategy we select a sample of products and query for $k=3$ and asked for our client to evaluate the answer. Based on this sample we realize the for distance higher than $0.4$ the products were completely different. 

So we come up with a strategy: first we check if the product has or not identifier, if has it goes to the normal pipeline, if not we need to generate the embeddings and them search the index with $k=1$ which is the closest neighbor, with that we will know the distance of this neighbor. 

\begin{figure}[htb]
\caption{Query Steps: (1) query point in blue; (2) select the closest neighbor $1-NN$, (3) Perform a range query to get all the similar points at the distance of closest neighbor; (4) retrieve all the points to compute price difference.}
\centering
\includegraphics[width=0.47\textwidth]{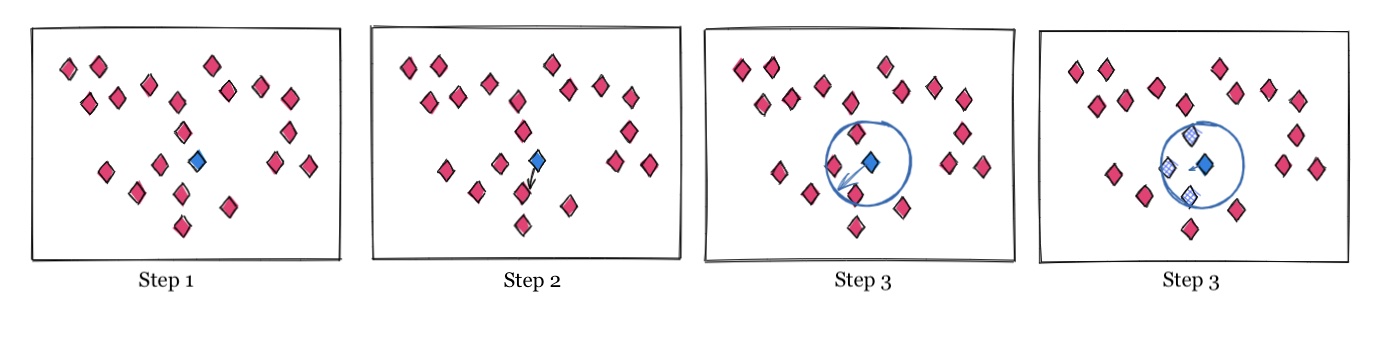}
\label{fig:querysteps}
\end{figure}

If the first neighbor is at most $0.4$ from the query point we add a value 0.00001 to the distance and perform a range query. This step will retrieve the tie list will all similar products that are in the same distance, we show this in Figure \ref{fig:querysteps}. 

\begin{figure}[h]
\caption{Full Pipeline showing the index creation part in black arrows and the query part in pink.}
\centering
\includegraphics[width=0.49\textwidth]{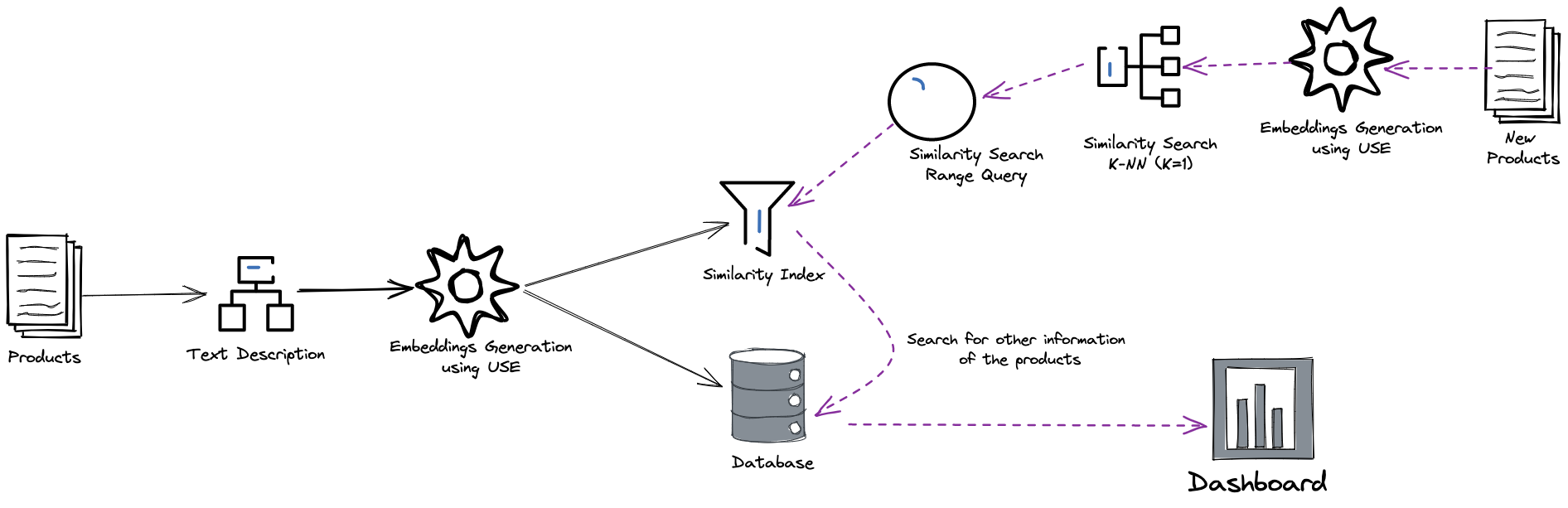}
\label{fig:pipeline2}
\end{figure}

After all candidates be retrieved we analyses other characteristic as price variations. 
For items that have distance higher than $0.4$ we flag as unique items. With all the matches done we save our results to a database and them to our dashboard to be evaluate by the final user. The query pipeline is presented in Figure \ref{fig:pipeline2}. 

This method was defined in collaboration with client teams and the information in our dashboard was defined after we did a shadow with one user that was the final user of the system.

\section{Results}

To test and validate our solution we run the analyses of three month of data, which corresponds to 72,175 thousand of items. 
From these purchases 43,562 were purchases with identifier and 28,613 without. In order to be fair we remove these orders from our index. 

\begin{table}[htb]
\caption{Distribution of analyzed items without identifier that match and consider unique.}
\begin{tabular}{|l|r|}\hline \hline
Result & No. items  \\ \hline \hline
Match & 15,866 \\ \hline
unique & 12,747  \\ \hline \hline
\end{tabular}
\label{tab:distribution}
\end{table}

From purchases without identifier we were able to match 55\% of the test and the rest was flagged as unique since the result has distance higher than $0.4$. We can see that a simple solution we are able to achieve a expressive result to expand the search for similar items. 

\begin{figure*}[htb]
\begin{minipage}{.48\textwidth}
\centering
\includegraphics[width=0.65\textwidth]{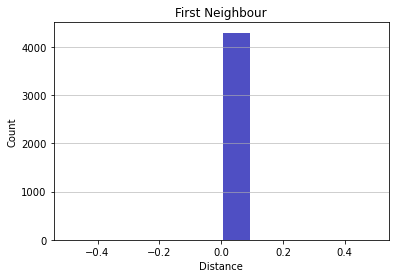}
\caption{(a) - Distance from the first neighbours - all points are in the index so the distance are zero.}
\label{fig:Dataset1a}
\end{minipage}%
\begin{minipage}{.48\textwidth}
\centering
\includegraphics[width=0.65\textwidth]{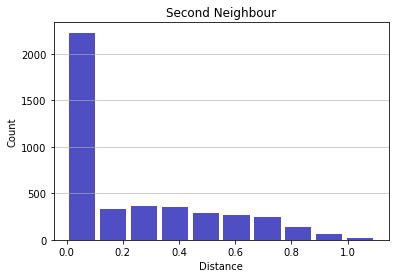}
\caption{(b) - Distance from the second neighbours - these are the real closest neighbour that are different from query point.}
\label{fig:Dataset1b}
\end{minipage}
\end{figure*}

\begin{figure*}[htb]
\begin{minipage}{.48\textwidth}
\centering
\includegraphics[width=0.65\textwidth]{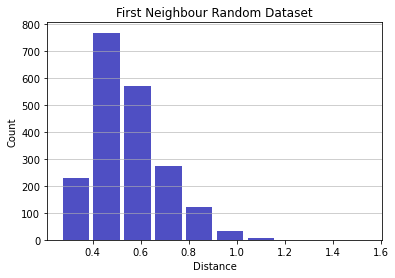}
\caption{(a) - Distance from the first neighbors - since all points are not in the index the distance is equal or higher 0.4.}
\label{fig:Dataset2a}
\end{minipage}%
\begin{minipage}{.48\textwidth}
\centering
\includegraphics[width=0.65\textwidth]{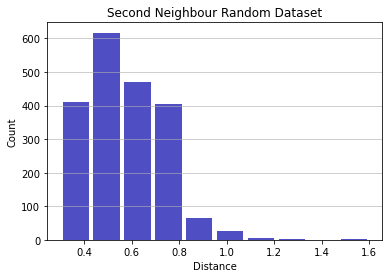}
\caption{(a) - Distance from the second neighbors - since all points are not in the index the distance is equal or higher 0.4.}
\label{fig:Dataset2b}
\end{minipage}
\end{figure*}

In order to show that the index was performing right and explain that we can only find if there is similar items we produce an experiment to explain how the index work and it is presented in Figures \ref{fig:Dataset1a}, \ref{fig:Dataset1b}, \ref{fig:Dataset2a} and \ref{fig:Dataset2b}. The questioning came about the items flagged as unique that are the ones with distance higher than $0.4$. 

For Figures \ref{fig:Dataset1a} and \ref{fig:Dataset1b} we select on month from the dataset and compute the K-NN search with $K=2$. Since we use data point that were indexed, the first neighbor should be the point itself and the distance should be zero. The second neighbor will be a different point and will show the distribution of distance of each neighbor. 

For Figures \ref{fig:Dataset2a} and \ref{fig:Dataset2b} we select 2000 examples from a dataset with Portuguese sentences used to training translation models and that is not the one used to create the index. Thus, the domain and the content of queries are completely different from the one used to create the index. As we see in Figure \ref{fig:Dataset2a} all the fist neighbors are more than $0.4$ from distance of query point. Which shows that they are very different. This proves that we cannot find in the index purchases that are not there. The Figure \ref{fig:Dataset2b} only reinforce the argument that the neighbors are very distant. 

Both experiments also shows the the choice of using only results that are up to $0.4$ from distance was a right choice.

\section{Conclusion}

This was a very challenging project, which was highly focus on data and we are able to delivery a MVP in 8 weeks that will be used right away what is exceptionally motivating and facilitates fast and successful business transformation for a data driven approach. 

This solution seems simple and shows that even a simple approach can provide the client the goal of find similar items. Our approach of work team up with the client allow a more assertive to solve the right problem and don't be lost in the way. Our user-centered methodology enable our solution assist the user perfectly. 

%%
%% The next two lines define the bibliography style to be used, and
%% the bibliography file.
\bibliographystyle{ACM-Reference-Format}
\bibliography{textbib}

\end{document}